\documentclass[a4paper]{article}
\usepackage{listings}
\usepackage{xcolor}
\usepackage{multirow}
\usepackage{jheppub}  
\usepackage{dcolumn}
\usepackage[english]{babel}
\usepackage[utf8x]{inputenc}
\usepackage[T1]{fontenc}
\usepackage{palatino}
\usepackage{amsmath}
\usepackage{afterpage}
\usepackage{graphicx, subcaption}
\usepackage[colorinlistoftodos]{todonotes}
\usepackage[colorlinks=true]{hyperref}
\usepackage{makeidx}
\newcommand{\be}{\begin{equation}}  
\newcommand{\ee}{\end{equation}}  
\newcommand{\bea}{\begin{eqnarray}}  
\newcommand{\eea}{\end{eqnarray}}  
\makeindex
\begin{document}

\vspace*{1.2cm}

\thispagestyle{empty}
\begin{center}
{\LARGE \bf PPS results and prospects from CMS/TOTEM collaborations}

\par\vspace*{7mm}\par

{

\bigskip

\large \bf Christophe Royon}

\bigskip

{\large \bf  E-Mail: christophe.royon@ku.edu}

\bigskip

{Department of Physics and Astronomy, The University of Kansas, Lawrence KS 66047, USA}

\bigskip

{\it Presented at the Workshop of QCD and Forward Physics at the EIC, the LHC, and Cosmic Ray Physics in Guanajuato, Mexico, November 18-21 2019}


\vspace*{15mm}

{  \bf  Abstract }

\end{center}
\vspace*{1mm}

\begin{abstract}
WWe describe the most recent results from the Proton Precision Spectrometer from the CMS and TOTEM collaborations, namely the first
observation of exclusive di-lepton production at high mass at the LHC,  and the prospects concerning the sensitivity to quartic anomalous couplings..
\end{abstract}

\section{Exclusive diffraction and photon-exchange}

\begin{figure}[h]
\centering
\includegraphics[width=1.in,clip]{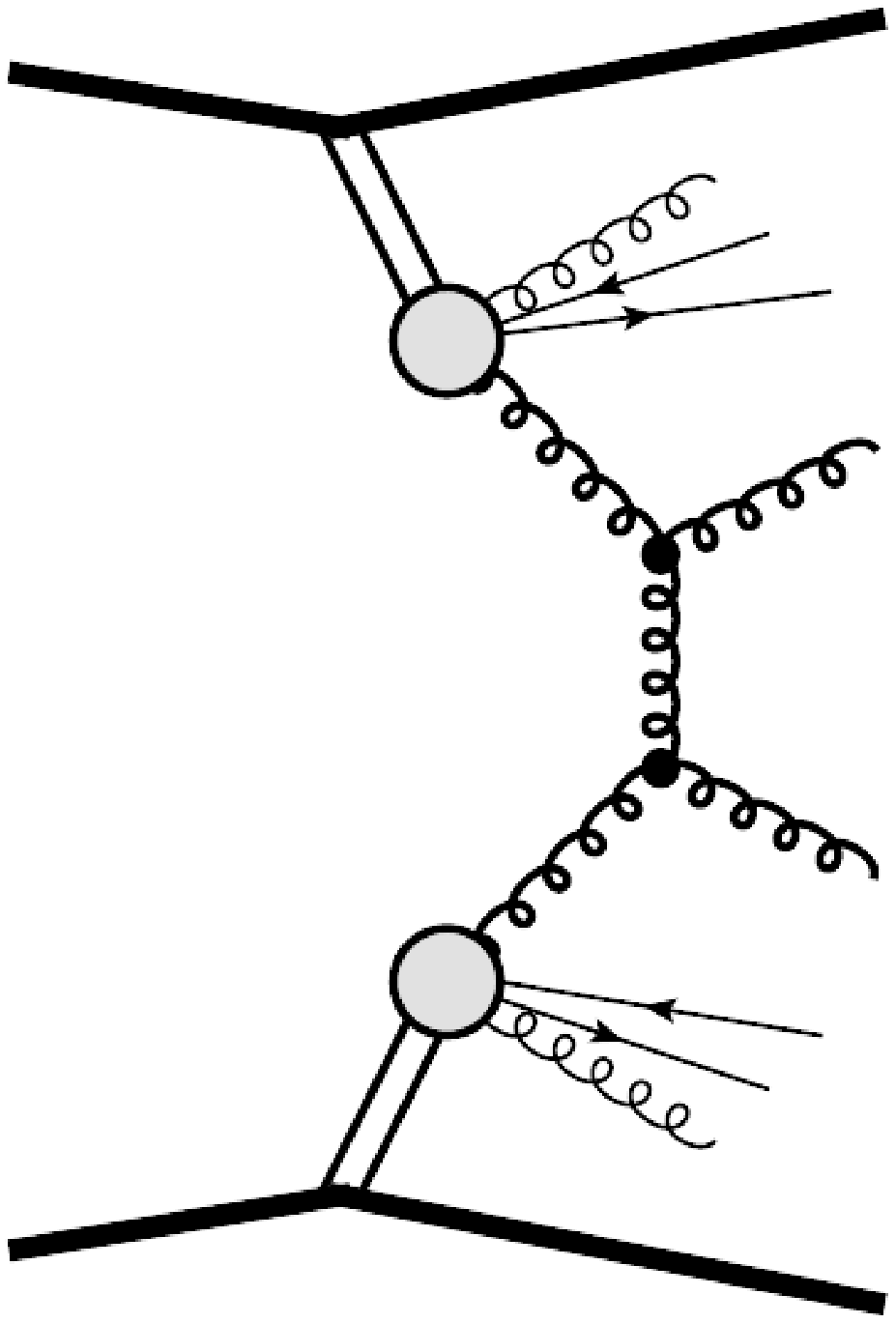}
\includegraphics[width=1.in,clip]{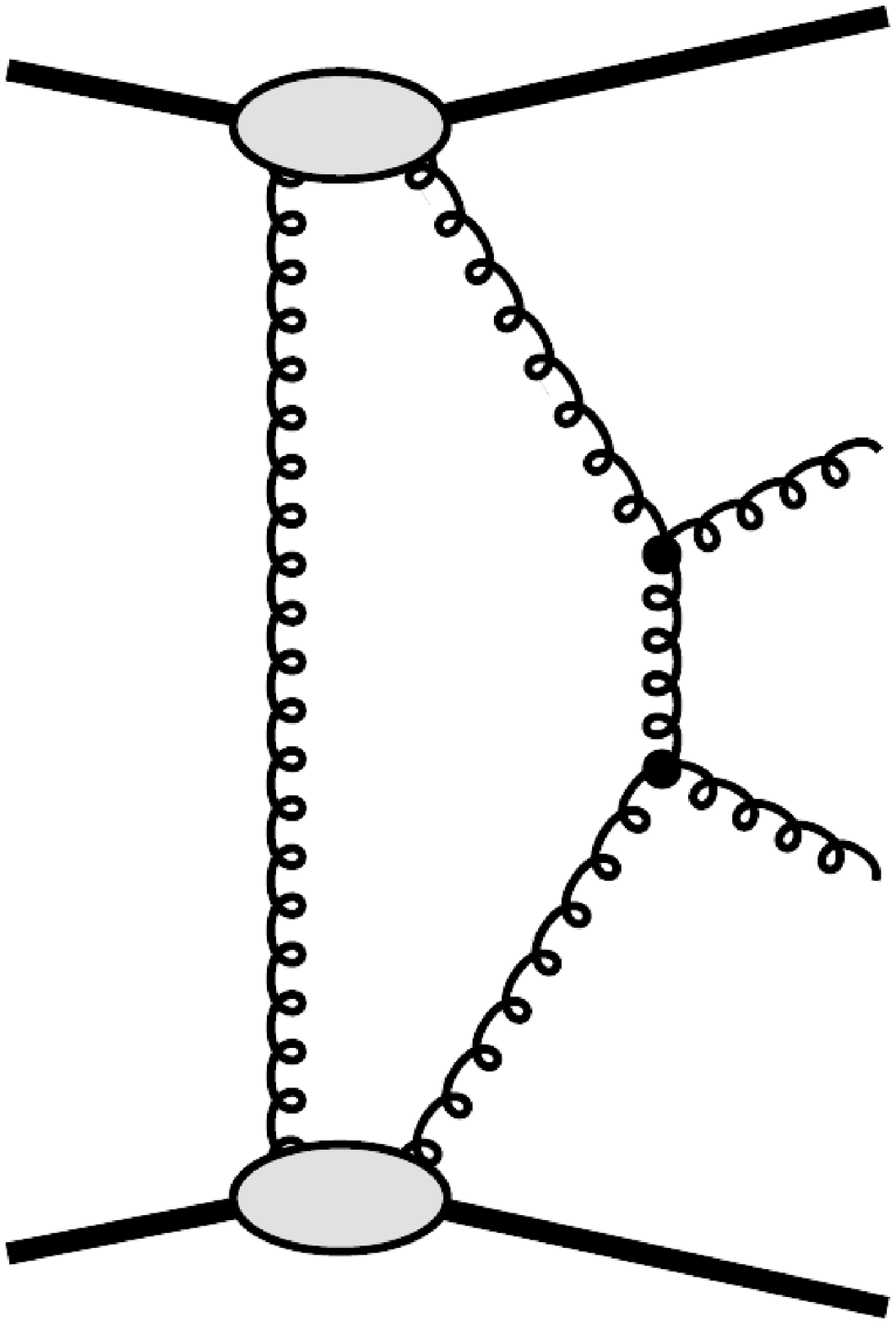}
\includegraphics[width=1.4in,clip]{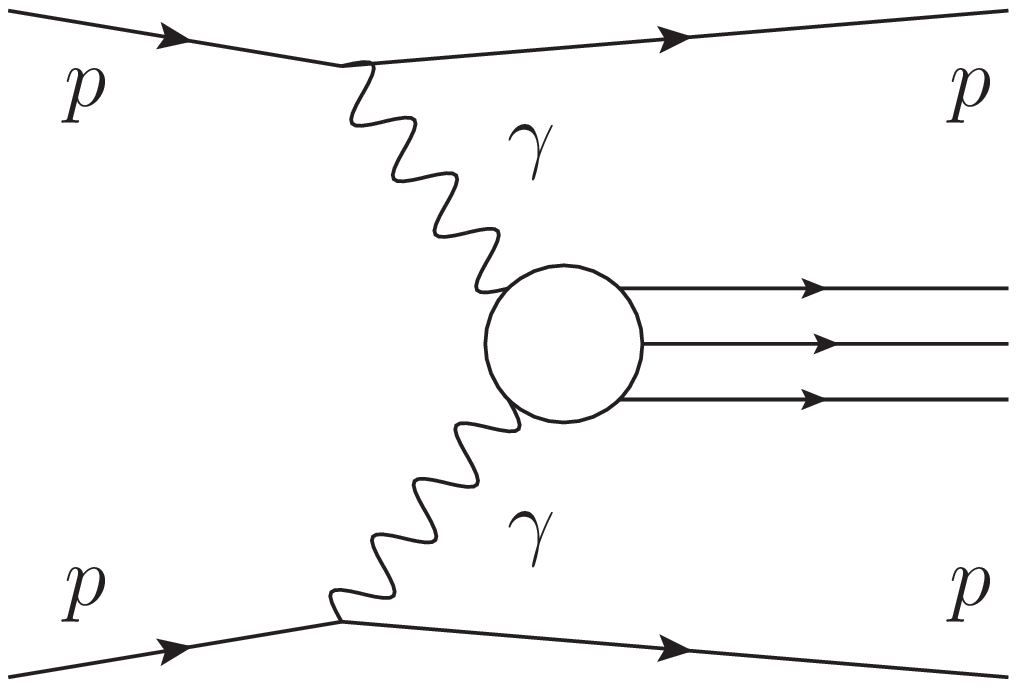}
\includegraphics[width=1.3in,clip]{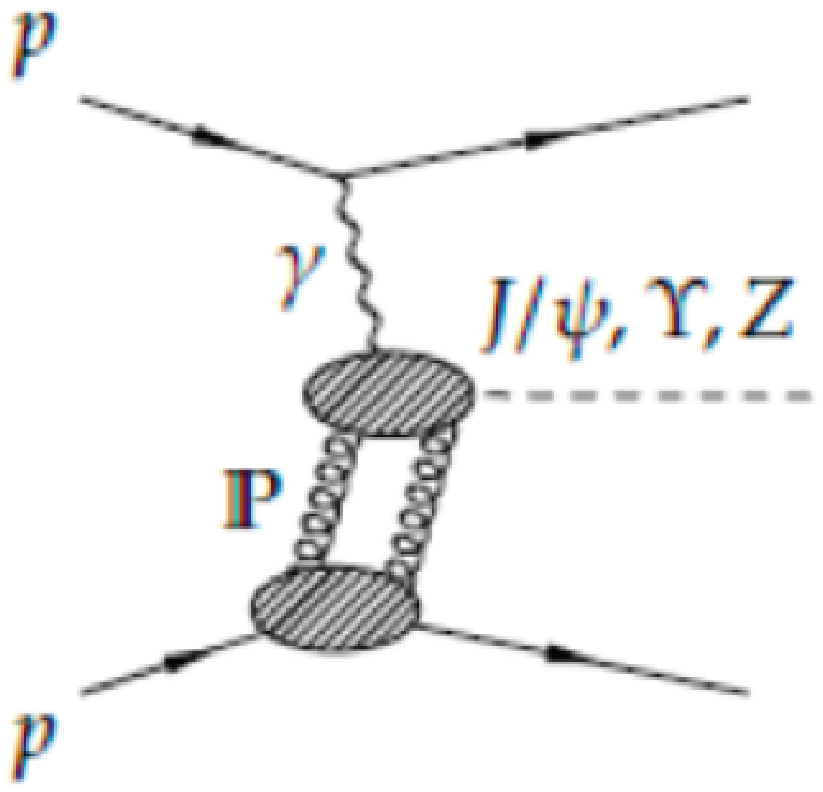}
\caption{Diagrams of inclusive and exclusive diffractive processes.}
\label{fig1}       
\end{figure}

We will first define what we call ``Exclusive" diffraction. The first left diagram of Fig.~\ref{fig1} corresponds
to Double Pomeron Exchange in inclusive diffraction. In this event, both protons are intact in the final state and two Pomerons
are exchanged. Gluons are extracted from each Pomeron in order to produce jets (or diphotons, $W$s...). Some energy is ``lost" in Pomeron remnants. The three 
other diagrams in Fig.~\ref{fig1} are exclusive in the sense that the full energy is used to produce the diffractive object. In other
word, there is no energy loss in Pomeron remnants. The second diagram corresponds to exclusive diffraction~\cite{KMR}, the third one
to photon exchanges and the last one to photon Pomeron exchanges that produce vector mesons. 
Exclusive events are specially interesting since it is possible to reconstruct the properties of the exclusively produced object from the tagged
proton. By comparing the information from the central CMS detector and the protons, it is possible to reduce the backgorund to a negligible
level~\cite{sylvain}.

\section{The Precision Proton Spectrometer}
\begin{figure}[h]
\centering
\includegraphics[width=5.in,clip]{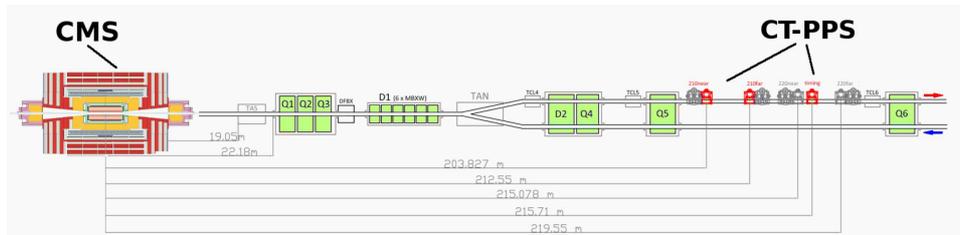}
\caption{Schematic view of the PPS detector by the CMS and TOTEM collaborations. The detectors are only depicted on one side of CMS for simplicity. Both sides of CMS are equipped with similar roman pot detectors. }
\label{pps}       
\end{figure}

The CMS-TOTEM Precision Proton Spectrometer~\cite{ctpps} (PPS)  was installed recently in order to detect and measure intact protons in the final state that lead to a possible better identification of exclusive events. The LHC  magnets bend the scattered protons outside the beam envelope. The roman pots detectors are located at about 210-220 m from the center part of the CMS  detector and cover a region in diffractive mass between typically 350 and 2000 GeV in standard luminosity runs at the LHC, depending on the exact beam lattice. A schematic view of the PPS roman pot detectors is shown in Fig.~\ref{pps}.
Two kinds of detectors are hosted in PPS.
The Silicon pixel and strip detectors allow to measure precisely the position of the intact proton and their distance from the beam axis, allowing 
to measure $\xi$, the proton fractional momentum loss, and $t$, the transferred energy squared at the proton vertex. In addition, timing 
detectors made of ultra-fast Silicon or diamond detectors allow measuring the proton time-of-flight.

The PPS detector started taking data in 2016 and could accumulate about 15 fb$^{-1}$ in 2016, and about 115 fb$^{-1}$ between 2016 and 2018. While the full data set is still being analyzed, we will describe the first observation of exclusive di-leptons using 9.4 fb$^{-1}$.  Roman pots are being inserted routinely during normal data taking at the LHC.

\begin{figure}[h]
\centering
\includegraphics[width=5.in,clip]{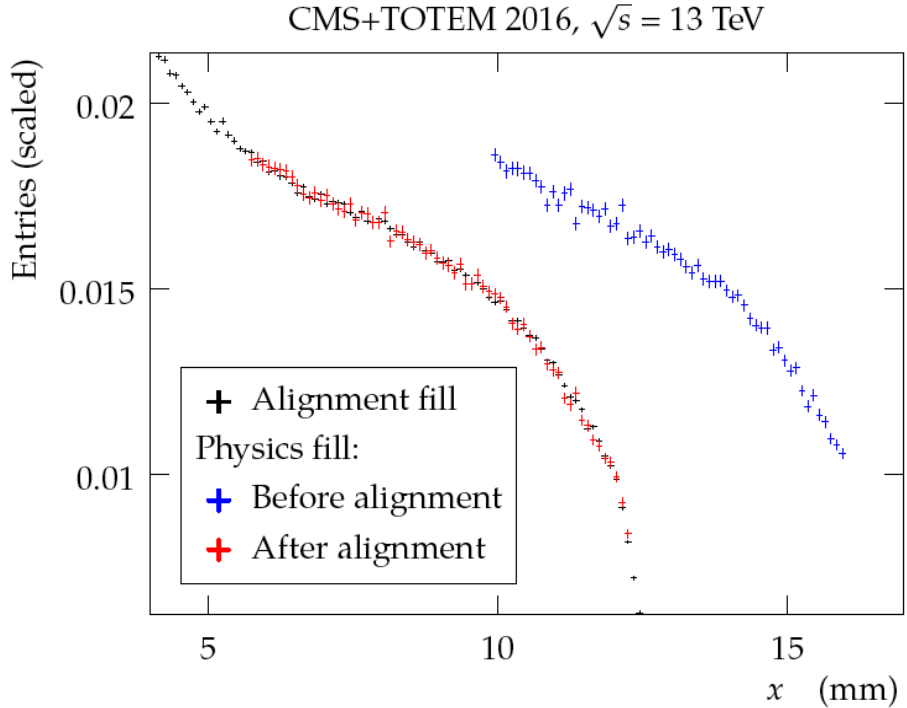}
\caption{Alignment procedure of the PPS detectors.}
\label{figalign}       
\end{figure}

One of the most difficult aspects of dealing with roman pot detectors is to align them with high precision
with respect to the beams. The procedure is described in Fig.~\ref{figalign}. The first step is the absolute alignment. 
The elastic $pp \rightarrow pp$ events in a special alignment run are used where both horizontal and vertical
roman pots get very close to the beam. This allows to obtain an absolute alignment of all vertical detectors with respect to the beam. The alignment of roman pots with respect to each other is then performed using inclusive events. This leads to the black points in Fig.~\ref{figalign}. The second step is to perform a relative alignment. The inclusive sample of protons triggered by CMS in standard runs is used and the distribution of proton track positions is matched to that of alignment as illustrated in Fig.~\ref{figalign} in blue and red points.

\section{Exclusive dilepton processes}

\begin{figure}[h]
\centering
\includegraphics[width=5.in,clip]{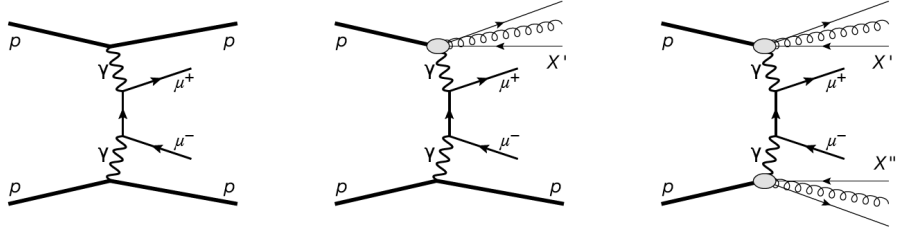}
\caption{Diagrams leading to di-muon production via photon exchanges.}
\label{fig8}       
\end{figure}

The CMS and TOTEM collaborations measured in 2016 the exclusive di-lepton production with 9.4 fb$^{-1}$. The corresponding diagrams are shown in Fig.~\ref{fig8}. The LHC is turned into a $\gamma \gamma$ collider and the flux of quasi-real photons is computed using the Equivalent Photon Approximation (EPA). 
Using the PPS detector, CMS and TOTEM  measured exclusive di-lepton production by tagging one intact proton in the final state. 
This is the first time the semi-exclusive di-lepton processes are measured with proton tag at high mass. In Fig.~\ref{fig8}, the two left diagrams correspond to the signal whereas the rightmost diagram is part of the background. The reason that only one proton is requested to be tagged is that less than one event is expected for double tagged events with about 10 fb$^{-1}$ of data due to the mass acceptance above about 350 GeV for the forward proton detectors. 

\begin{figure}[h]
\centering
\includegraphics[width=5.in,clip]{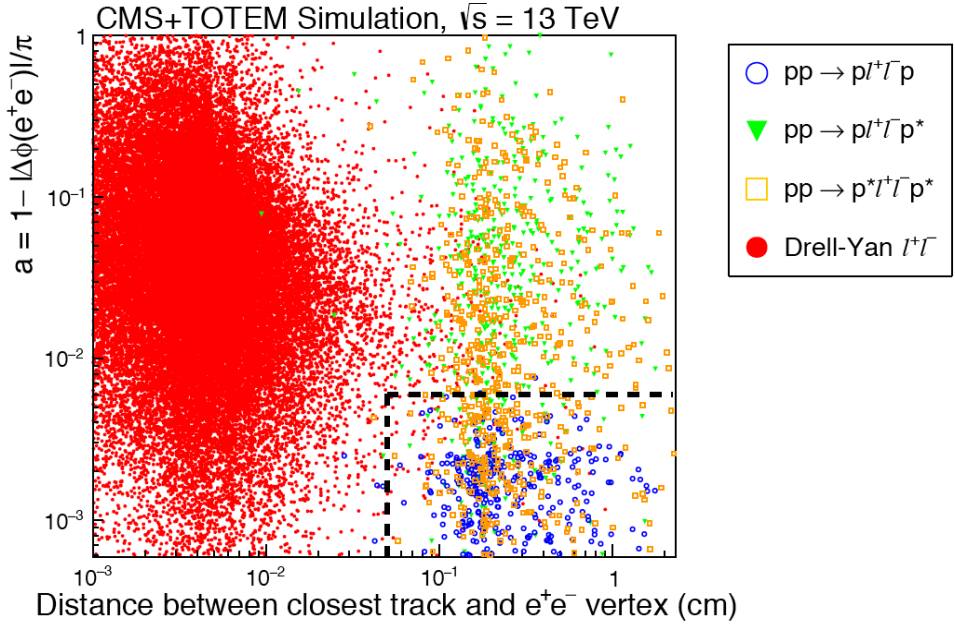}
\caption{Correlation between the acoplanarity and the distance between the closest track and the $e^+e^-$ vertex. The exclusive signal is in blue, while the single and double dissociative events are respectively in green and orange. The Drell-Yan background is in red. }
\label{veto}       
\end{figure}

\begin{figure}[h]
\centering
\includegraphics[width=5.in,clip]{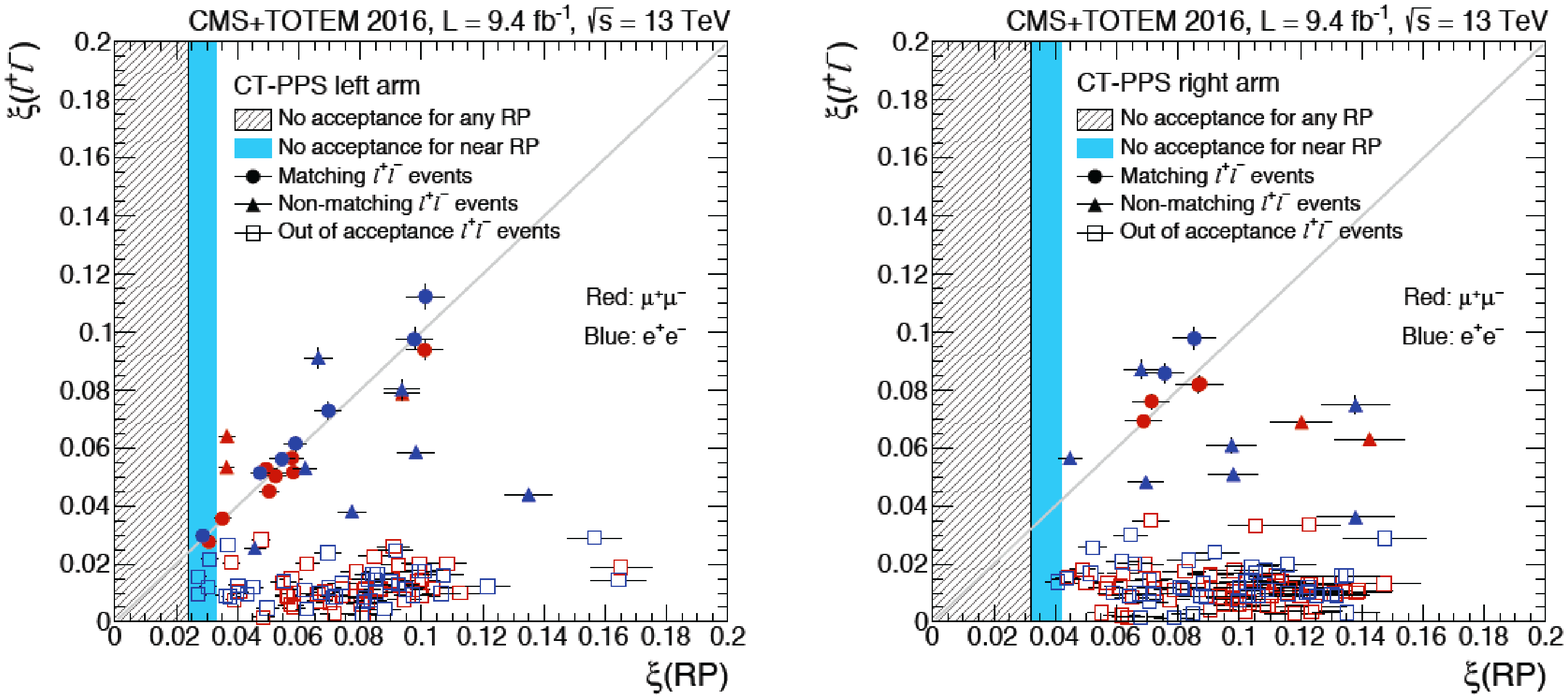}
\caption{Correlation between the $\xi$ values computed using the TOTEM roman pots and the di-lepton measured in CMS. The 20 semi-exclusive events are indicated in red. The left (right) plot displays the left (right) arm of TOTEM.}
\label{fig9}       
\end{figure}

A pair of opposite sign muons or electrons with $p_T>50$ GeV and $M_{l
l}>110$ GeV above the $Z$ boson peak is requested.
In order to suppress background, there is a veto on additional tracks around the di-lepton
vertex (within 0.5 mm) and leptons are required to be back-to-back, $|1 - \Delta
\Phi/\pi|<0.006$ for electrons (0.009 for muons) as shown in Fig.~\ref{veto}.
The main background is due to Drell-Yan di-lepton production with the intact proton originating from pile up events. This background is estimated using Drell-Yan $Z$ events in data and extrapolating from the $Z$ peak region to our exclusive di-lepton signal region. 40 events (17 $\mu \mu$ and 23 $ee$) are found with protons in the PPS acceptance and 20 (12 $\mu \mu$ and 8 $ee$) show a less than $2 \sigma$ matching between the values of $\xi$ computed using the TOTEM roman pots and using the di-lepton measured in CMS as shown in Fig.~\ref{fig9}~\cite{ctppsnote}. This leads to a significance larger than 5$\sigma$ to observe  20 events for a
background of $3.85$ ($1.49\pm 0.07 (stat) \pm 0.53 (syst)$ for $\mu \mu$ and 
$2.36\pm 0.09 (stat) \pm 0.47 (syst)$ for $ee$). As expected, no event was double tagged with an intact proton on each side.


\section{Anomalous coupling studies}

\begin{figure}[h]
\centering
\includegraphics[width=3.5in,clip]{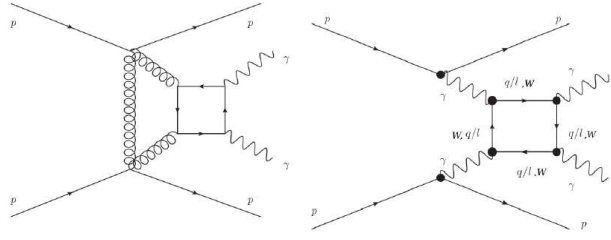}
\caption{Exclusive di-photon production. Left: QCD process, Right: QED process.}
\label{diagram}       
\end{figure}

\begin{figure}[h]
\centering
\includegraphics[width=3.5in,clip]{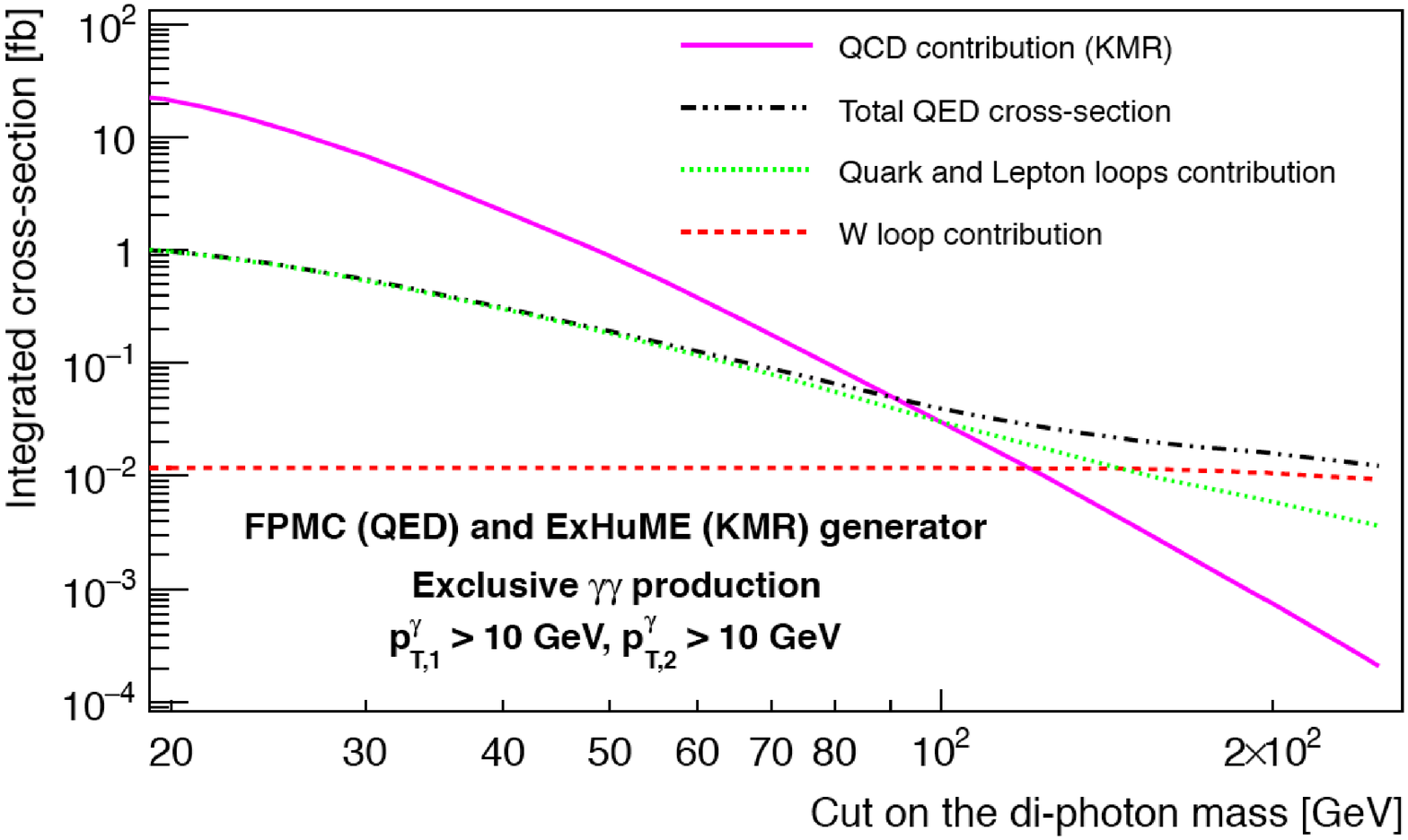}
\caption{Exclusive di-photon cross section above a given diphoton mass (in abscissa) for different processes.}
\label{smcross}       
\end{figure}

Using PPS, it is also possible to search for $\gamma \gamma \gamma \gamma$ anomalous couplings in a very clean way.
Within the acceptance of the PPS detectors during standard high luminosity runs at the LHC (basically for a di-photon mass above 350 GeV), it is possible to show that the exclusive production of di-photons is completely dominated by photon exchange processes and gluon exchanges can be neglected~\cite{anomalousgamma}. 
The QCD and QED diagrams leading to exclusive di-photon production are shown in Fig.~\ref{diagram}.
The di-photon exclusive cross sections are given in Fig.~\ref{smcross} for different contributions: QCD contribution in full purple line, QED contributions from quark and lepton loops in dashed green line, $W$ loop contribution in dotted red line, and the total QED contribution in black dashed dotted line. We note that the QCD contribution can
be neglected above a di-photon mass of 200 GeV. It means that measuring two photons in CMS and two protons in TOTEM corresponding to the same interactions  is a photon-induced process.

Four-photon couplings can be modified by loops of new particles or produced resonances that decay into
two photons. In case of loops $\zeta_1 = \alpha_{em}^2 Q^4 m^{-4} N c_{1,s} $ where the coupling depends only on the fourth power of the  charge and mass of the
charged particle, and on spin, $c_{1,s}$. This
leads to $\zeta_1$ of the order of 10$^{-14}$-10$^{-13}$. In case of resonances,
$\zeta_1 =
(f_s m)^{-2} d_{1,s}$ where $f_s$ is the $\gamma \gamma X$ 
coupling of the new particle to the
photon, and $d_{1,s}$ depends on the spin of the particle. For instance, 2 TeV
dilatons lead to $\zeta_1 \sim$ 10$^{-13}$.

\begin{figure}[h]
\centering
\includegraphics[width=3.5in,clip]{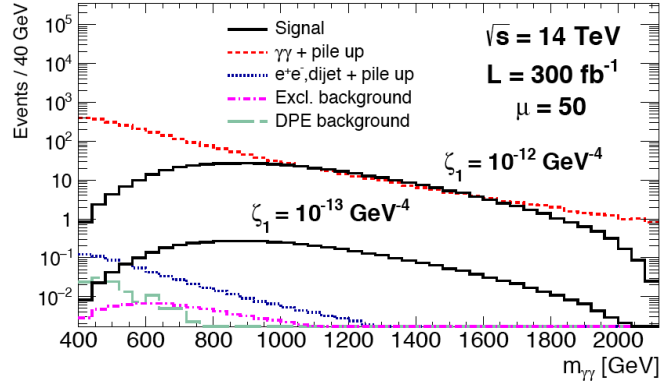}
\caption{Number of events as a function of the di-photon mass for signal ($\zeta_1=10^{-12}$ and $10^{-13}$ GeV$^{-4}$) and background.}
\label{cross}       
\end{figure}

The number of events for 300 fb$^{-1}$ as a function of di-photon mass is displayed in Fig.~\ref{cross} for signal
and background.  The exclusive di-photon and double Pomeron exchange (DPE) backgrounds are found
to be negligible at high mass. The only backgrounds that contribute at high mass are the non-diffractive di-photon
production $+$ pile up and di-lepton production $+$ pile up where leptons are misidentified as photons. Pile up 
events can be as large as 50 at the LHC at high luminosity and a typical pile up event contaminating our sample
will be made of one di-photon non-diffractive event and two intact protons originating from soft diffractive events.

\begin{figure}[h]
\centering
\includegraphics[width=2.9in,clip]{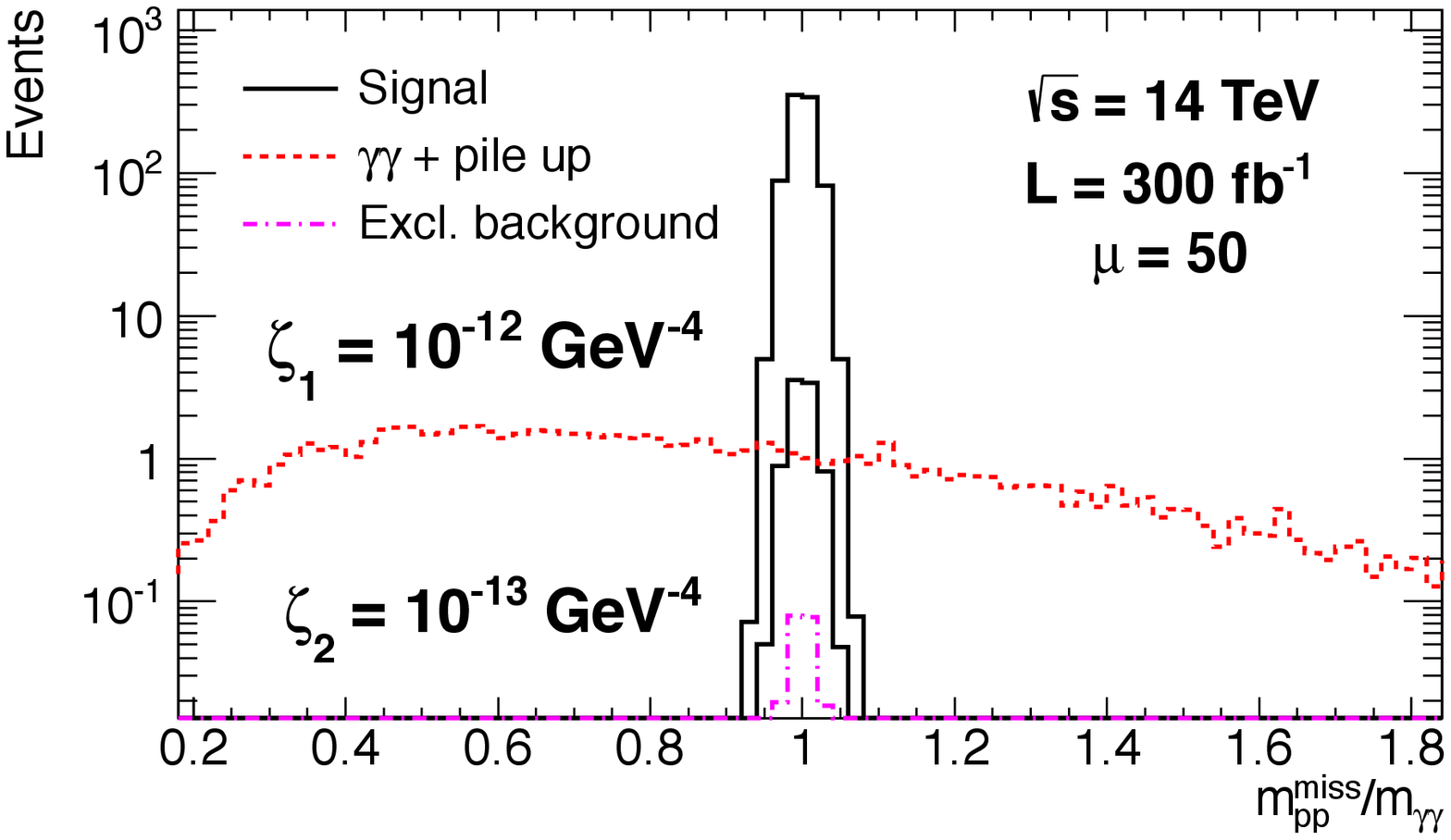}
\includegraphics[width=2.9in,clip]{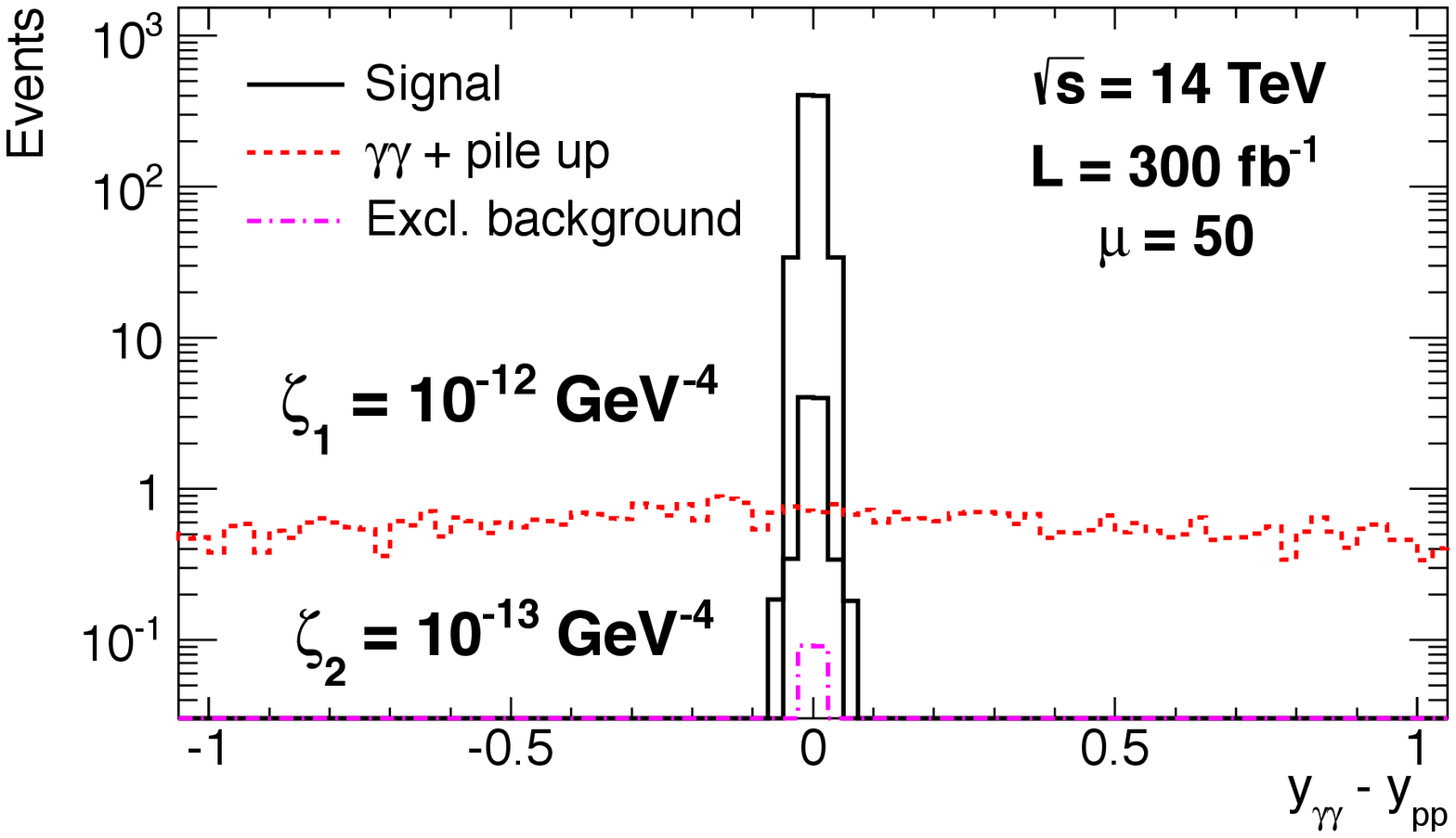}
\caption{Left: Ratio between the proton missing mass and di-photon mass for exclusive di-photon signal events and background. Right: Difference between the di-photon and di-proton rapidity for exclusive di-photon signal and background.}
\label{fig11}       
\end{figure}

\begin{figure}[h]
\centering
\includegraphics[width=4.8in,clip]{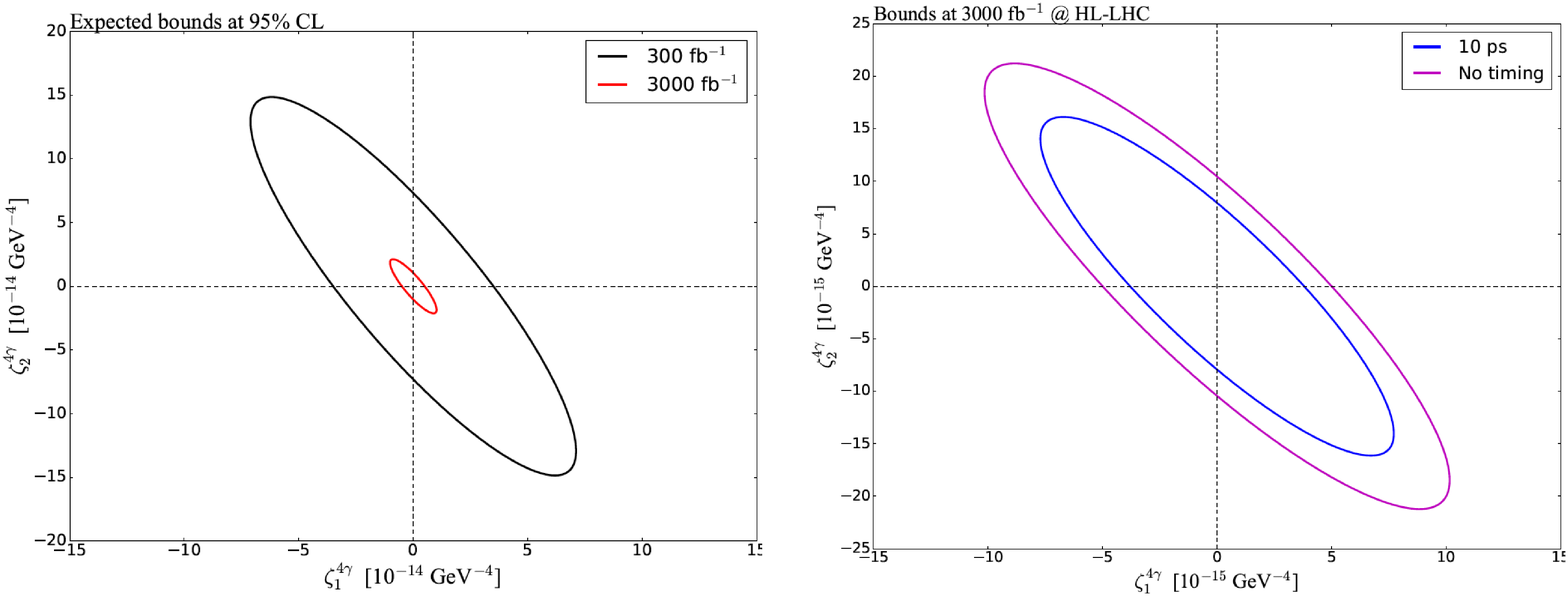}
\caption{Sensitivity contours on photon quartic anomalous couplings at the LHC with 300 and 3000 fb$^{-1}$.}
\label{exclusion}       
\end{figure}

Since the signal only shows two photons and two intact protons in the final state, we measure all final state particles. That allows us to obtain a negligible background for 300 fb$^{-1}$ at the LHC. The basic idea is to compare the proton missing mass and the di-photon mass as shown in Fig.~\ref{fig11}, left~\cite{anomalousgamma}. The signal peaks around 1.0 and the gaussian width is due to the detector resolution whereas the pile-up background leads to a much flatter distribution since the two protons are not related with the two photons. The same requirement can be performed using the difference in rapidity between the di-photon and di-proton systems, as shown in Fig.~\ref{fig11}, right. This leads to a background of less than 0.1 event for 300 fb$^{-1}$~\cite{anomalousgamma}. The gain on sensitivity compared to other methods at the LHC without detecting intact protons is more than two orders of magnitude on the $\gamma \gamma \gamma \gamma$ anomalous coupling, reaching values down to a few 10$^{-15}$.  It is worth noticing that, without exclusivity cuts described
in Fig.~\ref{fig11}, the background would be much larger for 300 fb$^{-1}$, namely about 80.3 events. We also 
extrapolated our results to high luminosity LHC for a luminosity of about 3000 fb$^{-1}$, and the sensitivity
can be even improved by a factor 10 as shown in Fig.~\ref{exclusion} in a conservative way. 

\begin{figure}[h]
\centering
\includegraphics[width=3.7in,clip]{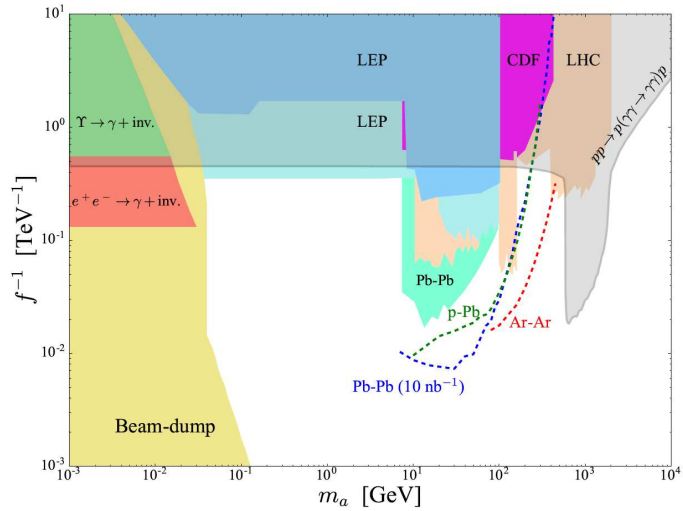}
\caption{Exclusion plot on axion like particles in the coupling versus mass plane and sensitivity at the LHC in $pp$ collisions with 300 fb$^{-1}$ (grey band) and in $PbPb$ 
(blue dashed line), $pPb$ (green dashed line), $ArAr$ (red dashed lines) collisions.}
\label{axion}       
\end{figure}

Looking for exclusive di-photon events with tagged protons can be directly applied to the search for axion-like particles (ALPs) at high mass~\cite{usaxion}. The ALP would be 
produced by $\gamma \gamma$ interactions and would decay into two photons. The sensitivity in the coupling versus ALP mass is shown in Fig.~\ref{axion} and we see the gain of about two orders of magnitude in coupling at high mass using this method with 300 fb$^{-1}$ at the LHC.
We also note that this is complementary to looking for exclusive di-photons in $pPb$, $PbPb$, and $ArAr$ collisions at lower masses of ALPs. This is due to the fact that the cross section is enhanced by a factor $Z^4$ in heavy ion collisions but the sensitivity at high mass is reduced to a large suppression at small impact parameter due to the size of the heavy ion~\cite{usaxionhin}.

The gain of two orders of magnitude on photon anomalous couplings is also true for $\gamma \gamma WW$ and $\gamma \gamma ZZ$  whereas the gain reaches three orders of magnitude for $\gamma \gamma \gamma Z$~\cite{oldaww,anomalousother}.  The search for anomalous couplings with tagged protons is now being pursued by the PPS collaboration.

\section{Conclusion}
In this report, we first described the first observation of high-mass 
exclusive dilepton production, leading to a  significance larger than 5$\sigma$ for observing 20 events for a
background of $3.85$ ($1.49\pm 0.07 (stat) \pm 0.53 (syst)$ for dimuons and 
$2.36\pm 0.09 (stat) \pm 0.47 (syst)$ for dielectrons. In a second part, we described the prospects for PPS,
leading to the  highest possible sensitivities
to $\gamma \gamma \gamma \gamma$, $\gamma \gamma WW$, $\gamma \gamma ZZ$,
$\gamma \gamma \gamma Z$ anomalous couplings that can appear due to new resonances, extra-dimensions. axion-like particles,
or composite Higgs... First results on these anomalous couplings are expected to come out soon.

\section*{Acknowledgements}

The author thanks the support from the Department of Energy, contract DE-SC0019389.



\end{document}